# A Foray into Efficient Mapping of Algorithms to Hardware Platforms on Heterogeneous Systems

Oren Segal, Nasibeh Nasiri, Martin Margala
Department of Electrical and Computer Engineering
University of Massachusetts Lowell
Lowell, USA
Oren_Segal/Nasibeh_Nasiri/Martin_Margala@uml.edu

*Abstract*— Heterogeneous computing can potentially offer significant performance and performance per watt improvements over homogeneous computing, but the question "what is the ideal mapping of algorithms to architectures?" remains an open one. In the past couple of years new types of computing devices such as FPGAs have come into general computing use. In this work we attempt to add to the body of scientific knowledge by comparing Kernel performance and performance per watt of seven key algorithms according to Berkley's dwarf taxonomy. We do so using the Rodinia benchmark suite on three different high-end hardware architecture representatives from the CPU, GPU and FPGA families. We find results that support some distinct mappings between the architecture and performance per watt. Perhaps the most interesting finding is that, for our specific hardware representatives, FPGAs should be considered as alternatives to GPUs and CPUs in several key algorithms: N-body simulations, dense linear algebra and structured grid. *(Abstract)*

## I. Introduction

In the last several years the idea that heterogeneous computing could offer performance and performance per watt improvements over homogeneous computing has gained roots in the engineering research community. But the question of which architecture is best suited for which type of algorithm remains open. Several benchmarks [1] [2] [3] [4] have been suggested in order to assess the merits of existing architectures. Perhaps the seminal work in this area was the Rodinia benchmark suite [2] which offered an initial way to standardize benchmarks for heterogeneous computing. In this work we investigate several algorithms on three different hardware architectures in an attempt to extract insights into a possible 'ideal' mapping between architecture and algorithm. Note that our findings are limited to the hardware we have access to and should not be taken as general conclusions regarding the relative merits of a specific architecture; instead these findings should be taken as a base for further investigation towards building a comprehensive database of algorithm implementations and their relative performance on different architectures. The main contribution of this paper is the added experimental data on the new and yet under-explored area of FPGA performance on high-level algorithms for HPC and general computing.

The rest of the paper is organized as follows: Section II describes the Rodinia benchmark suite. Section III describes our added Rodinia FPGA support. Section IV describes the hardware used in our experiments and the power usage model. Section V discusses the results. Section VI walks through an FPGA Kernel optimization example. Sections VII, VIII and IX are reserved for discussion, future directions and limitations of heterogeneous benchmarks respectively and finally section X is dedicated to related work.

## II. THE RODINIA BENCHMARK SUITE

Rodinia follows Berkeley's dwarf taxonomy [5]. The Berkeley's dwarf taxonomy is high level abstraction of common computing, memory access and communication patterns (Dwarves). It contains 13 classes and is based on a more limited 7 dwarves classification for scientific computing presented by Philip Colella in 2004 [6]. Rodinia creates implementations of algorithms that attempt to cover the range of algorithmic variance offered by the Berkeley's dwarf taxonomy. The available implementations are written in OpenCL [7], OpenMP and Cuda. Since we are interested in general heterogeneous computing not limited to GPUs or CPUs our focus is on the OpenCL implementations.

| Dwarf | Performance Limit: Memory Bandwidth, Memory Latency, or Computation? |
|---|---|
| 1. Dense Matrix | Computationally limited |
| 2. Sparse Matrix | Currently 50% computation, 50% memory BW |
| 3. Spectral (FFT) | Memory latency limited |
| 4. N-Body | Computationally limited |
| 5. Structured Grid | Currently more memory bandwidth limited |
| 6. Unstructured Grid | Memory latency limited |
| 7. MapReduce | Problem dependent |
| 8. Combinational Logic | CRC problems BW; crypto problems computationally limited |
| 9. Graph traversal | Memory latency limited |
| 10. Dynamic Programming | Memory latency limited |
| 11. Backtrack and Branch+Bound | ? |
| 12. Construct Graphical Models | ? |
| 13. Finite State Machine | Nothing helps! |

**Figure 1 - Dwarves Affinity assuming no bandwidth limitations[1]**

Assuming no limitation of bandwidth, Figure 1 summarizes the dwarves and their performance limits according to the authors of the Berkeley report.

---

[1] Taken from [5]

## III. FPGA SUPPORT

Rodinia is intended as a general benchmark suite for heterogeneous architectures but to the best of our knowledge was initially tested on CPUs and GPUs alone. In an effort to extend our knowledge to FPGA architectures, we made the necessary changes to Rodinia so it would be able to run on FPGAs. Altera OpenCL SDK (Versions 13/14) was used as the FPGA OpenCL implementation. At the time of conducting this research, Rodinia (Version 2.4) had a total of 19 benchmarks. One benchmark (MUMmerGPU) did not have an OpenCL implementation. Out of the available 18 OpenCL benchmarks, 15 compiled successfully and managed to run on Altera OpenCL for FPGAs. Two benchmarks worked only partially and two benchmarks could not fit inside the FPGA logic due to lack of sufficient logic elements. Future FPGAs, such as the Stratix 10 [8], with more logic elements, should be able to address this problem. The results confirm the general FPGA/OpenCL compatibility with the Rodinia set of benchmarks and by extension with the majority of the Berkley dwarves. In the next sections we will discuss the performance of seven of those benchmarks on different architectures.

## IV. HARDWARE EXPERIMENT DESIGN

We analyzed the performance of several of the Rodinia benchmarks on three different hardware architectures:

- Nallatech PCIe-385N Altera Stratix V Computing Card (8GB RAM, 150-250Mhz)
- Intel(R) Xeon(R) CPU L5630 @ 2.13GHz
- NVIDIA Tesla K40C

### A. Kernel Power Usage Model

**Table 1 Estimated Min and TDP Power Consumption**

| Device Name | TDP | Min Power Consumption |
|---|---|---|
| Nallatech PCIe-385N Altera Stratix V | 25W[2] | 19.5W[3] |
| Dual Intel(R) Xeon(R) CPU L5630 @ 2.13GHz | 80W[4] | 18.55W[5] |
| NVIDIA Tesla K40C | 235W[6] | 20.57W[7] |

Estimating Kernel power consumption is challenging since there is no standard way to measure kernel power across architectures, device power usage is dynamic, and it differs from algorithm to algorithm.

To accurately estimate the relative performance of the kernels running on the different architectures we used the power consumption data specified in Table 1. For each device we obtained the values either experimentally or through published manufacturer data. Thermal Design Power (TDP) [9] is the average power dissipation at maximum capacity. The Minimum Power Consumption (MPC) is either at device idle (FPGA/GPU) or least active state (CPU). Note that the TDP and the MPC are meant to serve as the upper and lower bounds of device power consumption while running a kernel. The significance of those numbers is that if a device A [kernel time * TDP] <= device B [kernel time * MPC] then device A is more power efficient than device B at running a specific kernel.

## V. RESULTS

In the following sections we will describe the individual performance and performance per watt of seven of the dwarves on a per algorithm basis. For each algorithm we report the kernel time and min/max power consumption range. Since an FPGA hardware design can be optimized according to an algorithm we report both the optimized and non-optimized performance results. Note that we do not change the algorithm in any way; instead we only use Altera OpenCL pragmas to optimize the design. In addition, when appropriate, we simplify math expressions whenever we believe the relatively young compiler can use the help. To distinguish between the two types of kernels the FPGA optimized version is marked (Opt) on the graphs. Note that we do not try to modify the original OpenCL code for the GPU/CPU. A review of the code suggests that it was already optimized towards GPU architectures by the original writers of Rodinia.

### A. FPGA Hardware Utilization Summary

**Table 2 Algorithm Hardware Utilization**

| Algorithm | Build Type | % Logic Elements | % Flip Flops | % RAMs | % DSPs | % Utilization |
|---|---|---|---|---|---|---|
| B+tree | Not-Opt-V13.1 | 15.7 | 13 | 32.5 | 0 | 27.6 |
| B+tree | Opt-V13.1 | 28.1 | 23.6 | 72.5 | 0 | 49.6 |
| BFS | Not-Opt-V13.1 | 15.4 | 11.5 | 25.4 | 0 | 25.6 |
| BFS | Opt-V13.1 | 46.5 | 32.6 | 85 | 0 | 75 |
| Hotspot | Not-Opt-V13.1 | 13.2 | 10.2 | 21.4 | 11.7 | 22.4 |
| Hotspot | Opt-V13.1 | 46.6 | 36.1 | 84 | 65.6 | 79 |
| Nearest Neighbor | Not-Opt-V13 | 10.5 | 7.3 | 14.5 | 1.6 | 17 |
| Nearest Neighbor | Opt-V13 | 45 | 32.4 | 60.1 | 50 | 73.8 |
| LavaMD | Not-Opt | 18.3 | 16.8 | 34.6 | 8.2 | 34.1 |
| LavaMD | Opt-V13 | 41.2 | 39.5 | 71.1 | 62.5 | 78.6 |
| Particle Filter | Not-Opt-V13 | 13.3 | 9.5 | 21.5 | 0 | 21.8 |

---

[2] Obtained from Nallatech published specifications [16] and verified by our experimental work [19]
[3] Obtained experimentally in our lab [19]
[4] Obtained from Intel Xeon Processor published specifications [18]
[5] Calculated from Intel Xeon Processor published specifications [18]; The power consumption represents the lowest active system state in which the dual Xeon L5630 is running at minimal core speed (1.6 GHz) and uses minimal voltage (0.75V).
[6] Obtained from NVIDIA published specifications [17]
[7] Obtained from NVIDIA published specifications [17]

| | | | | | | |
|---|---|---|---|---|---|---|
| Particle Filter | Opt-V14 | 38.2 | 27.6 | 84.1 | 0 | 62.5 |
| Stream Cluster | Not-Opt-V13.1 | 17.3 | 12.4 | 27.7 | 3.1 | 28.2 |
| Stream Cluster | Opt-V13.1 | 27.4 | 21.2 | 47 | 16.8 | 46.4 |

Table 2 summarizes the hardware utilization of the implemented algorithms. The Opt abbreviation stands for Optimized version. Higher utilization rates (with the exception of the BFS algorithm) translate to improved performance at the expense of higher usage of FPGA resources and total power, but the total FPGA power normally fluctuates between 19.5W (Idle) and 25W (active) regardless of the utilization rates. This can be explained by the fact that much of the design is shared among all kernels (PCIE/DDR, kernel infrastructure and interfaces etc.) and by the fact that there is currently no built-in power management interface for Altera OpenCL. This is in contrast to other devices such as CPUs/GPUs which can change their operating frequency and voltage, according to load, in order to conserve power.

### B. B+tree Algorithm

B+tree is a graph traversal search algorithm. It parallelizes tree search by assigning different search nodes to different threads. The kernel compute function is to compare key values [10]. As can be seen in Figure 2 the K40C GPU is 16.6X faster than the optimized FPGA version and that same optimized FPGA version is 2X faster than the CPU. Also note that the GPU is significantly more power efficient at B+tree than the other devices even at TDP levels. In addition it is highly likely that the FPGA is more power efficient than the CPU in this algorithm instance since the CPU will have to operate at near idle power range in order to compete with the FPGA. The hardware utilization for the B+tree FPGA implementation is shown in Table 2. As can be seen by the zero DSP utilization, this algorithm does not utilize any floating point hardware on the FPGA.

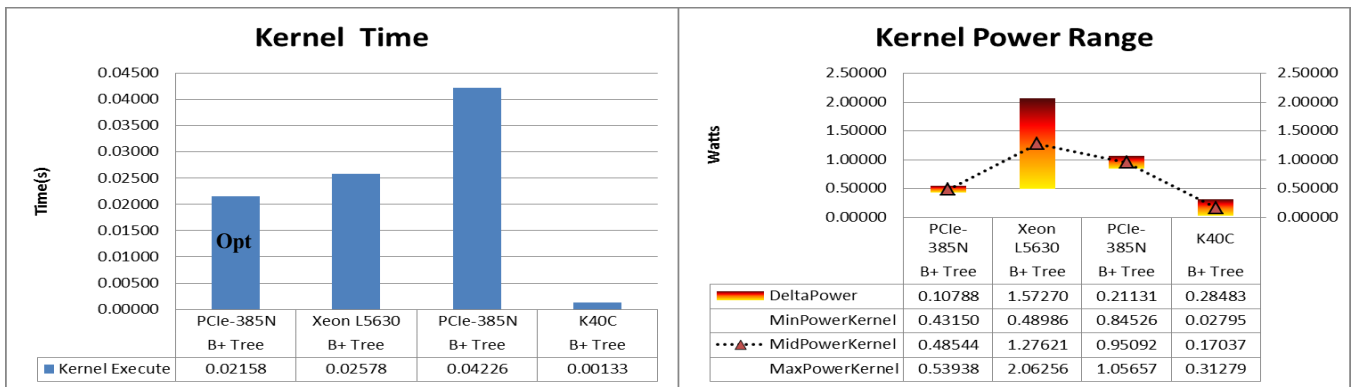

Figure 2 - B+tree

### C. BFS Algorithm

Breadth-First Search (BFS) is a graph search algorithm. It traverses all the connected components in a graph [2]. Both the Xeon Processor and the GPU perform significantly higher and are more power efficient than the FPGA implementation. Also note that this is the only benchmark in which the optimized version of the FPGA performs worse than the non-optimized version of the FPGA.

Table 2 shows the hardware usage of the kernels. Note that no DSPs are being used. This algorithm does not utilize any floating point hardware on the FPGA.

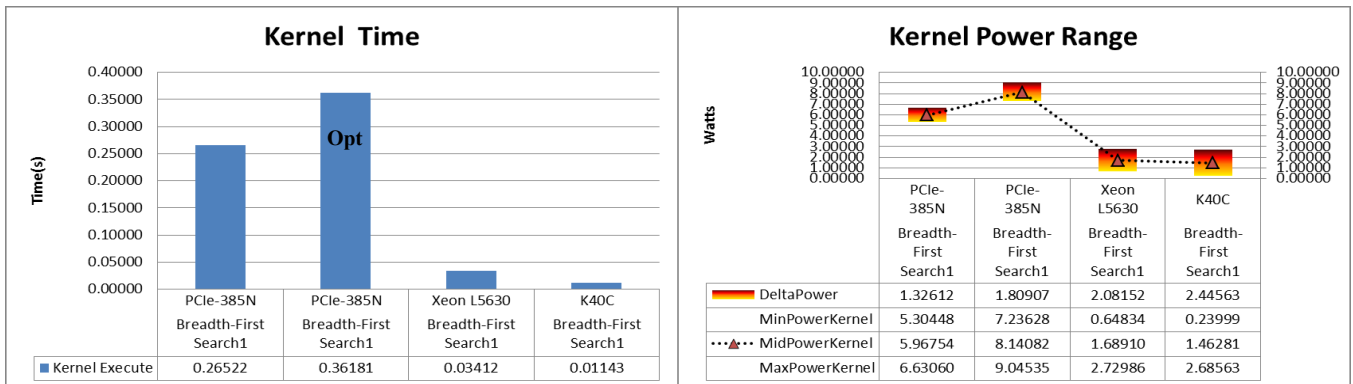

Figure 3 – BFS

## D. Hotspot Algorithm

Hotspot is classified as Structured Grid/Physics Simulation. It is a 2D transient thermal simulation kernel, which iteratively solves a series of differential equations for block temperatures.

As can be seen in Figure 4 the GPU dominates in performance and performance per watt in this algorithm implementation. Note that the performance of the non-optimized FPGA version is missing and not included. This algorithm makes use of the FPGA built in DSPs for floating point operations.

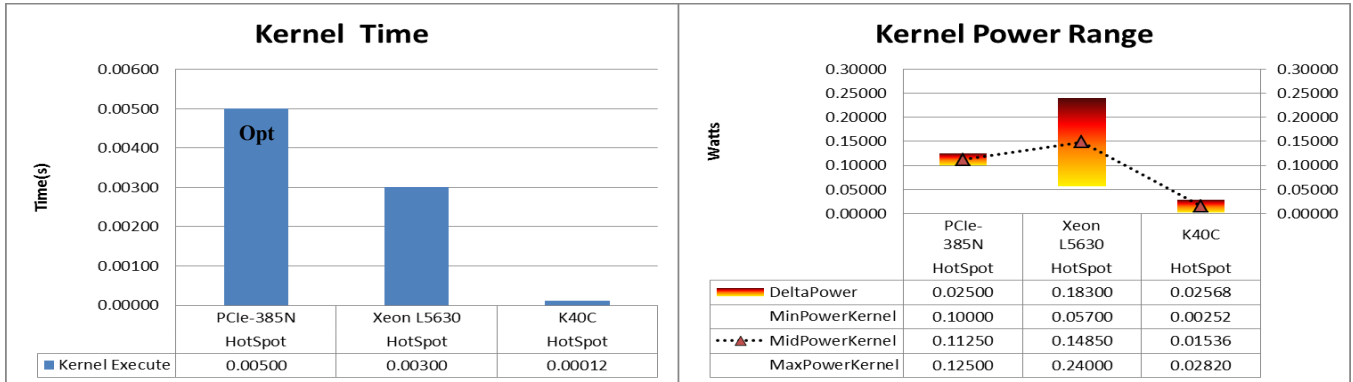

Figure 4 – Hotspot

## E. K-Nearest Neighbors Algorithm

K-Nearest Neighbors belongs to the dense linear algebra class. As can be seen in Figure 5 the FPGA kernel outperforms the Xeon CPU and is very close in performance to the K40C. If we look at the kernel power we can see that it is probably significantly more power efficient than both the CPU and the GPU.

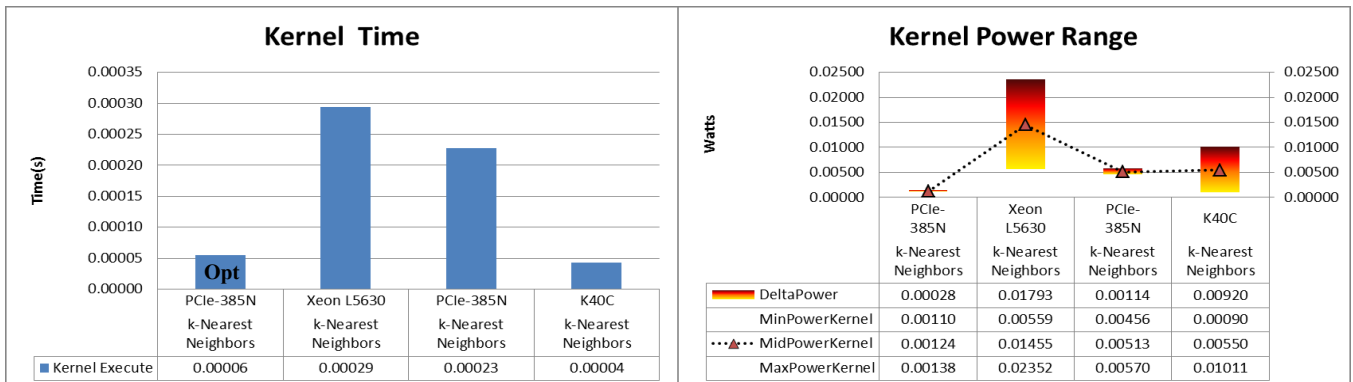

Figure 5 - KNN Results

## F. LavaMD Algorithm

LavaMD belongs to the N-body class. As can be seen in Figure 6 the performance of the optimized kernel is more than 4X faster than the CPU implementation. The GPU in turn is 3.2X faster than the FPGA version, but since the TDP of the FPGA is significantly lower than the GPU (25W vs 235W) it is very likely that the FPGA is more power efficient than the GPU in this type of kernel. The utilization rate of the LavaMD algorithm is relatively high and includes a high percentage of the DSPs on the FPGA chip.

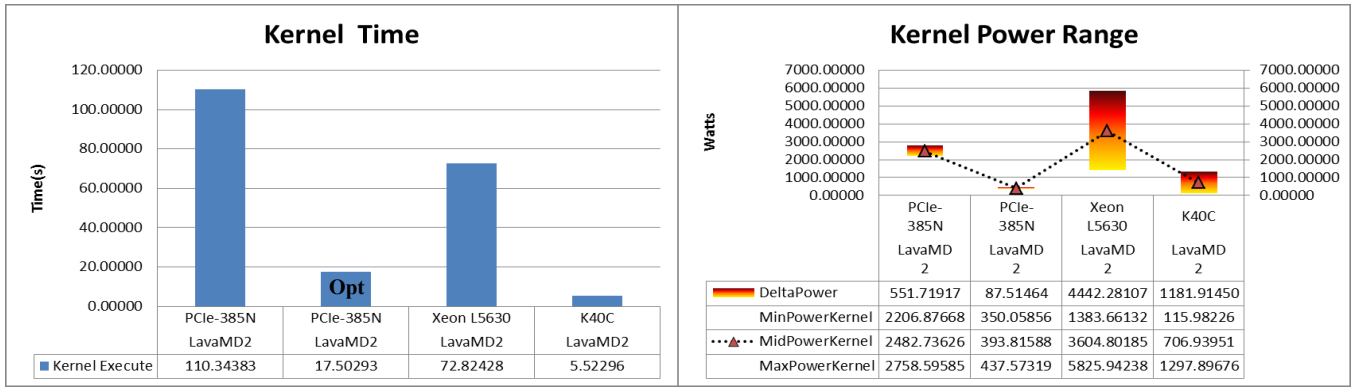

Figure 6 - LavaMD

### G. Particle Filter Algorithm

The Rodinia Particle filter is classified as a structured grid algorithm. It comes in three different flavors: naïve, single precision and double precision. The only version we managed to compile and run on the FPGA was the naïve version. Figure 7 shows the performance results of the naïve version. It is very likely that the FPGA is more power efficient than the CPU and GPU. The naïve version of the algorithm does not use DSP components on the FPGA.

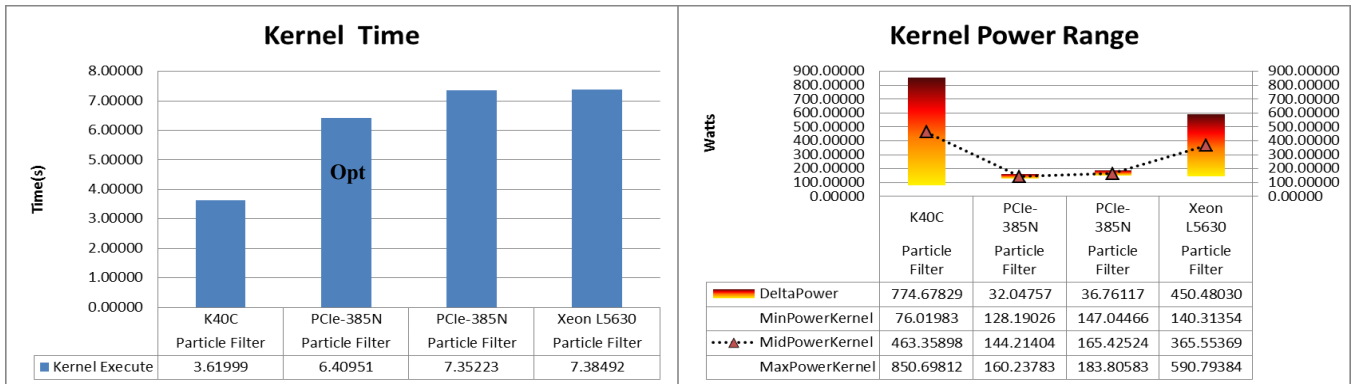

Figure 7 - Particle Filter

### H. Stream Cluster Algorithm

Stream cluster belongs to the Dense Linear class (Algebra Data Mining domain). The GPU dominates in performance and performance per watt. The hardware utilization is less than fifty percent even in the optimized version of the FPGA algorithm, suggesting that it is not an ideal algorithm implementation for an FPGA.

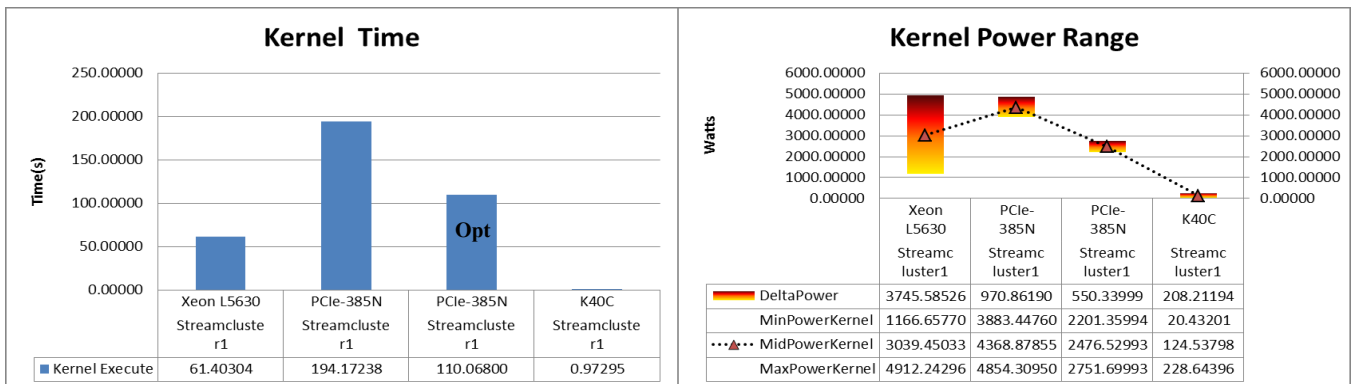

Figure 8 – Streamcluster

## VI. KERNEL OPTIMIZATION EXAMPLE

Figure 9 shows a kernel optimization example. We use the original Nearest Neighbor Rodinia kernel as the base code. We then do the following:

1. Add/set Altera OpenCL specific directives (num_compute_units, restrict, loop_unroll etc.)
2. Simplify expressions – although they should be done automatically by the compiler, we find that it helps the compiler produce more optimized code if we simplify expressions manually
3. Finally we compile the code using auto optimizations (O3)

The parts marked in red are the parts that have been modified by us. The resulting Kernel is 3.8X faster than the non-optimized kernel.

Note that the optimization process is based on intuition and trial and error, for example to find the optimal number of compute units we experimented with several versions of the kernel with varying pragma variations: num_compute_units(16), num_compute_units(8) etc. Unfortunately it is currently not practical to do an exhaustive search of all options because of compilation time. We hope that future improvements in compilation techniques will reduce compilation time from many hours for a highly optimized design to several minutes or preferably several seconds.

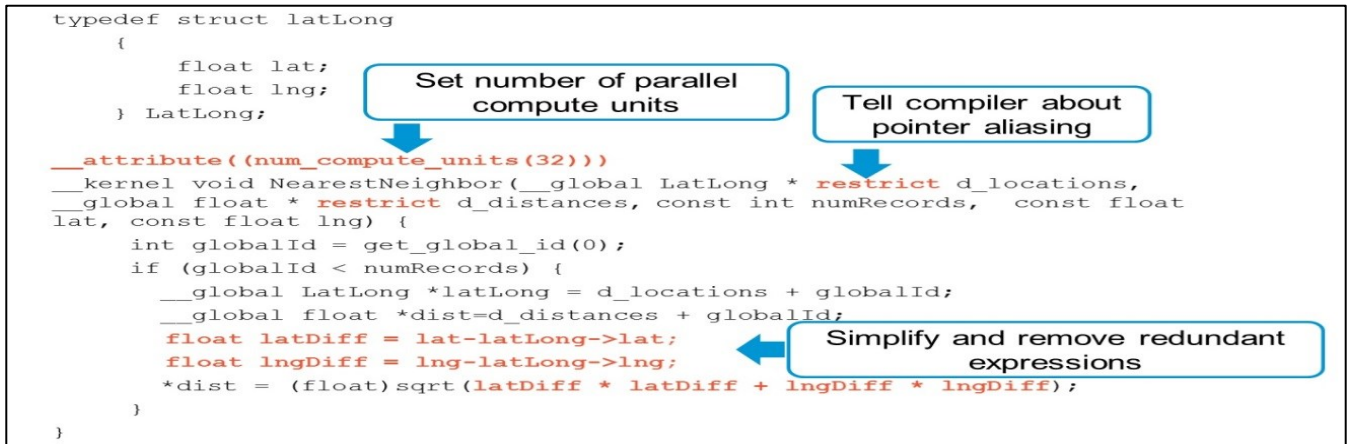

**Figure 9 – Nearest Neighbor Kernel Optimization Example**

## VII. DISCUSSION

While it is difficult to generalize from the results on different architecture merits, the results do seem to support the idea that in some areas FGPAs should be considered as alternatives to CPUs/GPUs. Research towards specialized implementations of these algorithms on FPGAs might lead to implementations that yield even higher performance and or performance per watt. If we follow the Berkeley classification then N-body simulations, dense linear algebra and structured grid should be investigated first.

The FPGA device, in its current Altera OpenCL implementations (V13.0-15.1), has very stable power consumption which only varies by several watts (19.5-25W) during execution of different kernels. On the other hand the CPU and GPU employ sophisticated power management techniques and power consumption can vary significantly according to algorithm and resource usage. For example, during our experiments, the K40C would move between 30 and 230 watt power consumption in a matter of seconds. If FPGAs are to be efficiently utilized in data centers, research into FPGA power management should be considered a priority since idle time is a part of life in such environments [11].

Two more critical aspects for overall performance of accelerators are off-chip bandwidth [12] and the energy cost of moving data off-chip [13]. In this work we analyze kernel performance/performance per watt alone, but it is important to mention that many of the performance benefits of accelerators can be lost when hardware architectures force moving data off-chip between accelerators and/or CPU. In the end the relationship between compute intensity, memory bandwidth and energy costs can dictate the viability of using an accelerator to perform an algorithm even if the accelerator kernel itself is faster or more power efficient than another CPU or accelerator type.

## VIII. FUTURE DIRECTIONS

Testing more of the Rodinia benchmarks on additional types of FGPAs/GPUs/CPUs/DSPs and future hardware would hopefully reveal more interesting leads and venues for future potential hardware exploration. We believe that for heterogeneous benchmarks to be meaningful we need to go beyond CPUs and GPUs and build a comprehensive database of results across multiple hardware types and algorithms. Such an endeavor is crucial for the understanding of the relative merits of different architectures using heterogeneous benchmark suites.

## IX. LIMITATIONS OF STANDARD BENCHMARKS

While it is important to run heterogeneous benchmarks on different architectures we should also remember their limitations. If the benchmarks are to be used as guidelines to hardware manufacturers developing new hardware and to compiler developers then they offer some important hints but they risk creation of specialized "benchmark purpose" hardware which is of limited use in real world scenarios [14].

In addition, if predictions are correct then we are at the beginning of a new exascale era in which heterogeneity will play a major roll. In this era we will have to deal with combinations of factors that will make static benchmarking of limited use. Instead of the one accelerator scenario that we typically have now, we could be looking at multiple types of accelerators running in tandem. In such a case we would need to be able to benchmark the combined performance of the accelerators to achieve optimal performance. This combined performance is not the same as individual device performance. The performance instead would depend on interactions between various components in the system and the communication-bus architecture. In addition, other seemingly unrelated system components such as active cooling components will affect the overall performance per watt and should be considered in order to achieve optimal performance per watt.

Such complex interactions would be difficult to model and anticipate. It is more likely that accurate benchmarking would only be attainable on a per system-instance basis and should be dynamically evaluated continually since the rate of failure in exascale systems would be high [13] and each failure would essentially create a new "machine type" with different types of available resources.

## X. Related Work

An inclusive cross hardware heterogeneous benchmark suite needs to support a common language base. Directive based languages such as OpenACC limit control and hardware-specific languages such as CUDA are not suitable for direct cross hardware comparison. The current open standard for parallel computing is the Open Computing Language (OpenCL). Several high level heterogeneous benchmarks [1] [2] [3] [4] based on OpenCL exist, but to the best of our knowledge optimization and reported results are provided for CPUs and GPUs alone. An FPGA specific benchmark called CHO has recently been proposed [15]. It offers benchmarking OpenCL for FGPAs as an alternative to traditional HLS benchmarks. CHO includes several arithmetic, media and cryptographic algorithms but does not include the rich set of high-level algorithms that Rodinia has to offer and does not attempt to cover the Berkeley's dwarf taxonomy.

## Acknowledgment

We would like to thank Altera, NVidia, and Nallatech for their hardware donations so that we can develop and test the framework on different architectures. We would also like to thank HP for supporting various stages of this research over the past several years.

## Acknowledgment

We would like to thank Altera, NVidia, and Nallatech for their hardware donations so that we can develop and test the framework on different architectures. We would also like to thank HP for supporting various stages of this research over the past several years.